\documentclass[%
 reprint,
 superscriptaddress,
 amsmath,amssymb,
 aps,
]{revtex4-1}

\usepackage{graphicx}
\usepackage{xspace}
\usepackage{color}
\usepackage{url}
\usepackage{dcolumn}
\usepackage{bm}

\usepackage{setspace}

\begin{document}

\preprint{APS/123-QED}

\newcommand{\mug}{$(g_{\mu}-2)$}
\newcommand\rev[1]{#1}
\newcommand\reva[1]{#1}
\newcommand{\Mum}{$\mathrm{Mu^{-}}$}
\newcommand{\mup}{$\mu^{+}$}

\title{\reva{Compact} buncher cavity for muons accelerated by a radio-frequency quadrupole}
\newcommand{\KEK}{\affiliation{High Energy Accelerator Research Organization (KEK), Tsukuba, Ibaraki 305-0801, Japan}}
\newcommand{\JAEA}{\affiliation{Japan Atomic Energy Agency (JAEA), Tokai, Naka, Ibaraki 319-1195, Japan}}
\newcommand{\JPARC}{\affiliation{J-PARC Center, Tokai, Naka, Ibaraki 319-1195, Japan}}
\newcommand{\Tokyo}{\affiliation{University of Tokyo, Hongo, Tokyo 171-8501, Japan}}
\newcommand{\Ibaraki}{\affiliation{Ibaraki University, Mito, Ibaraki 310-8512, Japan}}
\newcommand{\Nagoya}{\affiliation{Nagoya University, Nagoya, Aichi 464-8602, Japan}}
\newcommand{\Riken}{\affiliation{Riken, Wako, Saitama 351-0198, Japan}}
\newcommand{\SNU}{\affiliation{Seoul National University, Seoul 08826, Republic of Korea}}
\newcommand{\INPA}{\affiliation{Institute for Nuclear and Particle Astrophysics, Seoul National University, Seoul, 08826, Republic of Korea}}
\newcommand{\KU}{\affiliation{Korea University, Seoul 02841, Republic of Korea}}
\newcommand{\BINP}{\affiliation{Budker Institute of Nuclear Physics, SB RAS, Novosibirsk 630090, Russia}}
\newcommand{\NSU}{\affiliation{Novosibirsk State University, Novosibirsk 630090, Russia}}
\newcommand{\Pulkovo}{\affiliation{Pulkovo Observatory, St. Petersburg, 196140, Russia}}

\author{M. Otani}\email{masashio@post.kek.jp}\KEK
\author{Y. Sue}\email{ysue@hepl.phys.nagoya-u.ac.jp}\Nagoya


\author{K. Futatsukawa}\KEK


\author{T. Iijima}\Nagoya

\author{H. Iinuma}\Ibaraki



\author{N. Kawamura}\KEK


\author{R. Kitamura}\JAEA


\author{Y. Kondo}\JAEA 


\author{T. Morishita}\JAEA

\author{Y. Nakazawa}\Ibaraki

\author{H. Yasuda}\Tokyo
\author{M. Yotsuzuka}\Nagoya


\author{N. Saito}\JPARC



\author{T. Yamazaki}\KEK


\begin{abstract}
A buncher cavity has been developed for the muons accelerated by a radio-frequency quadrupole linac (RFQ). 
The buncher cavity is designed for $\beta=v/c=0.04$ at an operational frequency of 324~MHz. 
It employs a double-gap structure operated in the TEM mode for the required effective voltage with compact dimensions, 
in order to account for the limited space of the experiment. 
The measured resonant frequency and unloaded quality factor are 323.95~MHz and $3.06\times10^3$, respectively. 
The buncher cavity was successfully operated for longitudinal bunch size measurement of the muons accelerated by the RFQ. 


\end{abstract}

\maketitle


\section{Introduction}
Muon linacs have been studied for their potential advantages in various branches of science. 
After the muons are cooled to thermal energy~\cite{bib:muonium1, bib:muonium2}, the muons are accelerated to the specific energy required by an applications. 
One of the applications of accelerated muon beams is the transmission muon microscope~\cite{bib:mutrans}, which is used in materials and life sciences. 
If the muons are accelerated up to 10~MeV, it enables three-dimensional imaging of living cells, which is impossible with the use of transmission electron microscope. 
Another application of the muon linac is the precise measurement of the muon anomalous magnetic moment \mug\ and electric dipole moment at the Japan Proton Accelerator Research Complex (J-PARC E34)~\cite{bib:e34}. 
In the J-PARC E34 experiment, the muons are accelerated to 212~MeV by a series of acceleration cavities~\cite{bib:murfq1, bib:murfq2, bib:ih, bib:daw, bib:dls}. 
Recently muon acceleration to 89~keV using a radio-frequency quadrupole linac (RFQ)~\cite{bib:ptorotype_rfq} was demonstrated~\cite{bib:murfq}. 
In order to reach the specific energy required for various applications, it is necessary to perform additional acceleration with successive RF cavities. 
Beam matching between the RF cavities is extremely important, as is the case with the linac for proton or ions, in order to avoid substantial emittance growth. 
Especially in the muon linac for the J-PARC E34 experiment, the longitudinal beam matching after the RFQ using buncher cavities is extremely important 
because a cavity followed by the RFQ adopts an alternative phase focusing (APF) scheme~\cite{bib:apf1, bib:apf2}; 
there is a strong correlation between the longitudinal and transverse directions in the APF scheme, and the longitudinal mismatch results in emittance growth in both the transverse and longitudinal directions. 
\par
A buncher cavity has been developed for the muon accelerated by the RFQ. 
The designed particle velocity $\beta$ is 0.04, and the operational frequency is 324~MHz, which is the same as that of the RFQ. 
From among the different types of room temperature cavities, a quater-wave resonator with a double-gap is employed to account for the limited space of the experiment. 
The buncher cavity for the accelerated muons presented here has a unique feature; it has a compact structure compared to the buncher cavities developed for protons or heavy ions~\cite{bib:SNS_mebt, bib:SNS_mebt2, bib:C_ADS, bib:j-parc}. 
\rev{The results presented in this paper are first demonstration of the buncher cavity dedicated for muons, 
which is essential step to realize a new-generation muon beam discussed for several applications~\cite{bib:mulinac1, bib:mulinac2, bib:mulinac3}. }
\reva{The buncher cavity presented here is an exotic one by reason of its compactness. 
It will stimulate new opportunities in the field. } 
\par
In this paper, we describe the design and results of the buncher cavity. 
In Sec.~\ref{sec:design}, the designs and fabrications are described. 
Section \ref{sec:result} is devoted to the results obtained in the RF measurements and operation. 
The conclusions are presented in Sec.~\ref{sec:conclusion}
\section{Design and fabrication}\label{sec:design}
The buncher cavity is designed for measuring the longitudinal bunch size after acceleration with the RFQ. 
The experiment was conducted at the J-PARC muon science facility (MUSE)~\cite{bib:muse}. 
The J-PARC MUSE provides a pulsed muon (\mup) beam with a 25-Hz repetition rate. 
The muons are decelerated by an aluminum degrader, and some portions form negative muonium (\Mum, $\mu^{+}\ e^{-}\ e^{-}$). 
The \Mum's are extracted and accelerated to 5.6~keV by an electrostatic lens~\cite{bib:soa}. 
They are then injected into the RFQ and accelerated to 89~keV. 
Then, the accelerated \Mum's are transported to a detector via a diagnostic beamline. 
\par
\begin{figure}[!htbp]
   \begin{center}
    \includegraphics[width=0.5\textwidth]{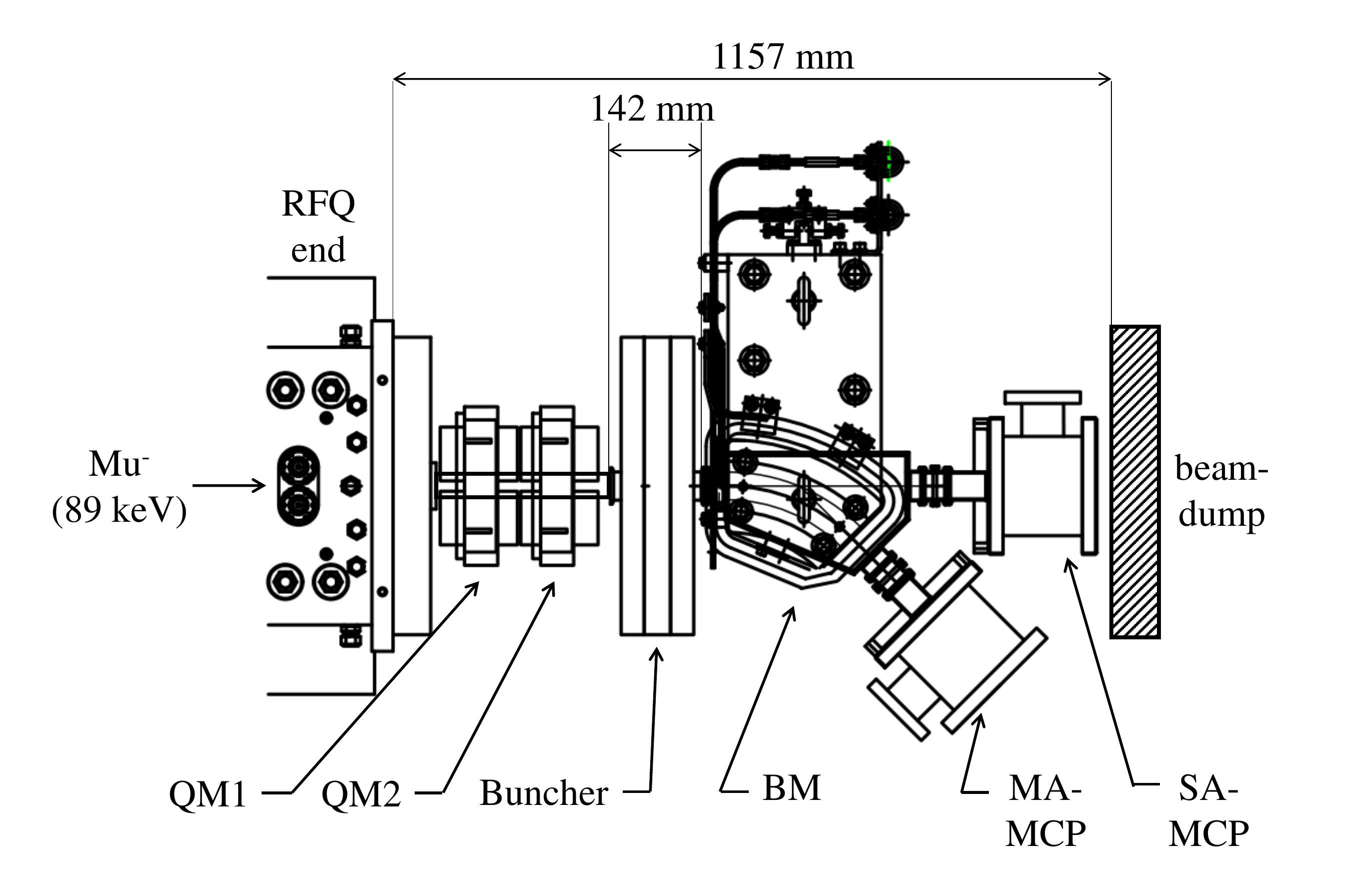}
   \end{center}
   \caption{Layout of the diagnostic beamline for the muon acceleration experiment using the RFQ. } 
   \label{fig:layout}
\end{figure}

Figure~\ref{fig:layout} shows the layout of the diagnostic beamline. 
In previous experiments for the demonstration of muon acceleration~\cite{bib:murfq} and profile measurement~\cite{bib:murfq_profile}, 
the diagnostic beamline consisted of a pair of quadrupole magnets (QM1 and QM2) and a bending magnet (BM). 
In this experiment for longitudinal bunch size measurement, a buncher cavity is added 
between the quadrupole magnet and the bending magnet. 
A single-anode (SA) microchannel plate (MCP, Hamamatsu  photonics F9892-21~\cite{bib:hamamatsu_mcp}) and multi-anode (MA) MCP detectors are placed at the downstream end of the straight line and $45^{\circ}$ line, respectively. 
The SA-MCP detector measures the penetrating $\mu^{+}$~\cite{bib:murfq} for the beamline tuning. 
The MA-MCP detector is used for the longitudinal bunch size measurement~\cite{bib:det}. 
In order to focus the beam longitudinally, and measure the longitudinal beam properties, it is necessary to have a buncher cavity. 
Because there is a beam-dump just after the detectors, the available space for the buncher cavity is approximately 150~mm. 
\par
\par
The function of the buncher cavity is to provide sufficiently large effective voltage for the longitudinal bunch size measurement 
with the available space. 
In order to estimate the required effective voltage, a series of simulations are performed. 
The muon beamline is simulated using g4beamline~\cite{bib:g4bl}. 
The conversion from \mup\ to \Mum\ is implemented using the data from a separate experiment~\cite{bib:mum_exp}. 
The simulation of the electrostatic lens was conducted using GEANT4~\cite{bib:g4} \reva{in order to estimate not only \Mum\ but also positrons generated from decay of muon stopped in the electrostatic lens. }
In the simulation, the electric field of the electrostatic lens was calculated using OPERA~\cite{bib:opera}. 
PARMTEQM~\cite{bib:parmteqm} was employed for the RFQ simulation \reva{because it can reproduce data well from the experience at the J-PARC linac. }
For the end cell section, PIC simulation with GPT~\cite{bib:GPT}, in which the electric field calculated with 
CST MW Studio~\cite{bib:cst} was implemented, was employed to estimate the effects due to 
the unequal spacing of the vanes at the end. 
TRACE3D~\cite{bib:trace3d} \reva{was used to design the beam optics} and PARMILA~\cite{bib:parmila} \reva{was employed to obtain particle distribution} for the
diagnostic beamline simulation. 
\par
\begin{figure}[!htbp]
   \begin{center}
    \includegraphics[width=0.5\textwidth]{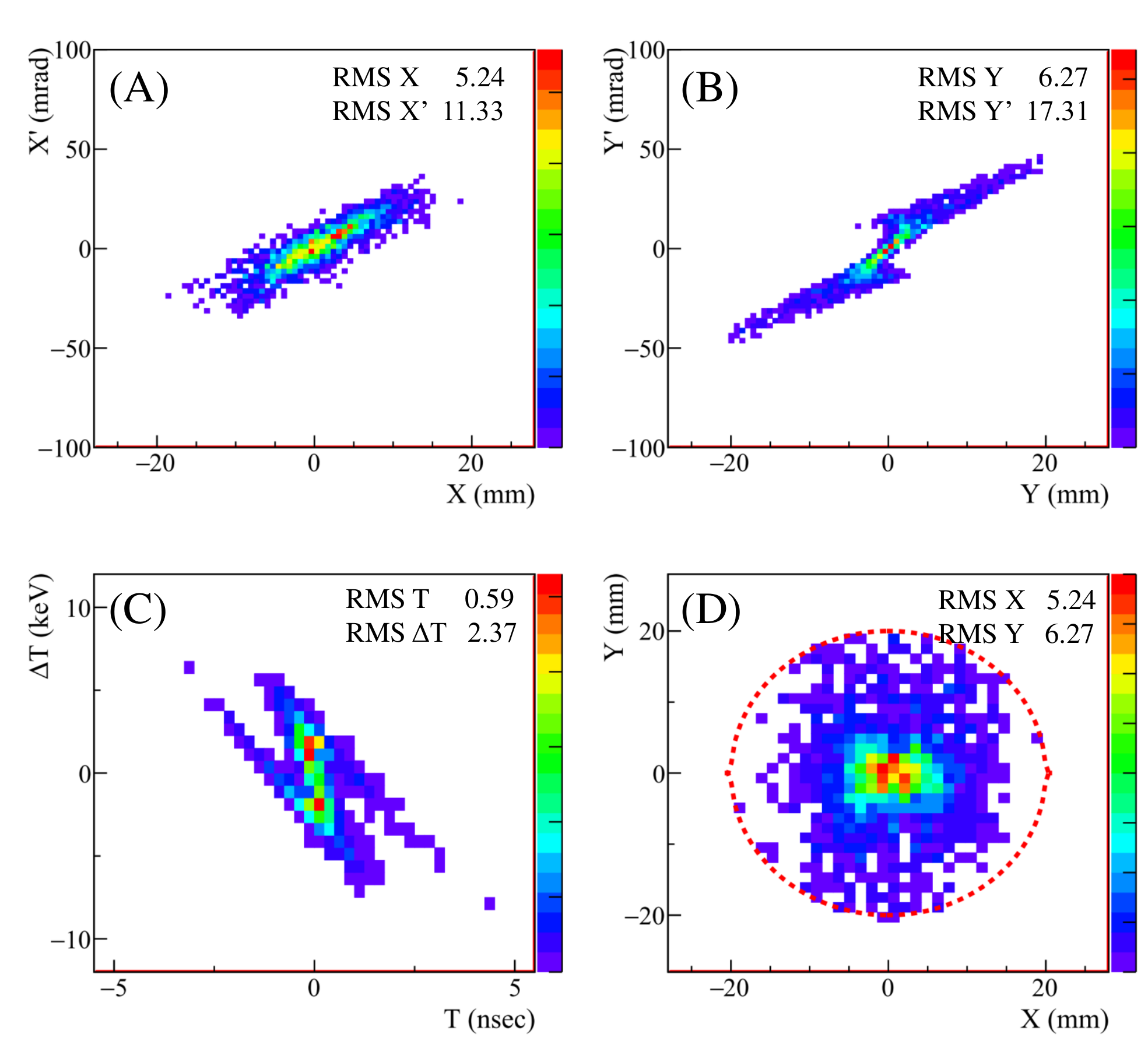}
   \end{center}
   \caption{Calculated phase space distributions at the MA-MCP detector location. (A) the horizontal divergence angle $x'$ vs $x$, 
   (B) the vertical divergence angle $y'$ vs y, (C) $\Delta W$ vs $\Delta t$, and (D) y vs x. The red dotted circle in (D) represents the effective area of the MA-MCP detector. }
   \label{fig:t3d}
\end{figure}

Figure~\ref{fig:t3d} shows the simulated phase-space distributions at the MA-MCP detector location. 
The longitudinal bunch width is estimated to be 600~psec in rms, which is sufficiently small compared to the detector sensitivity~\cite{bib:det}. 
On the basis of the simulaton, the required effective voltage for the buncher is estimated to be 5.3~kV. 
\par
\par
\begin{figure}[!htbp]
   \begin{center}
    \includegraphics[width=0.45\textwidth]{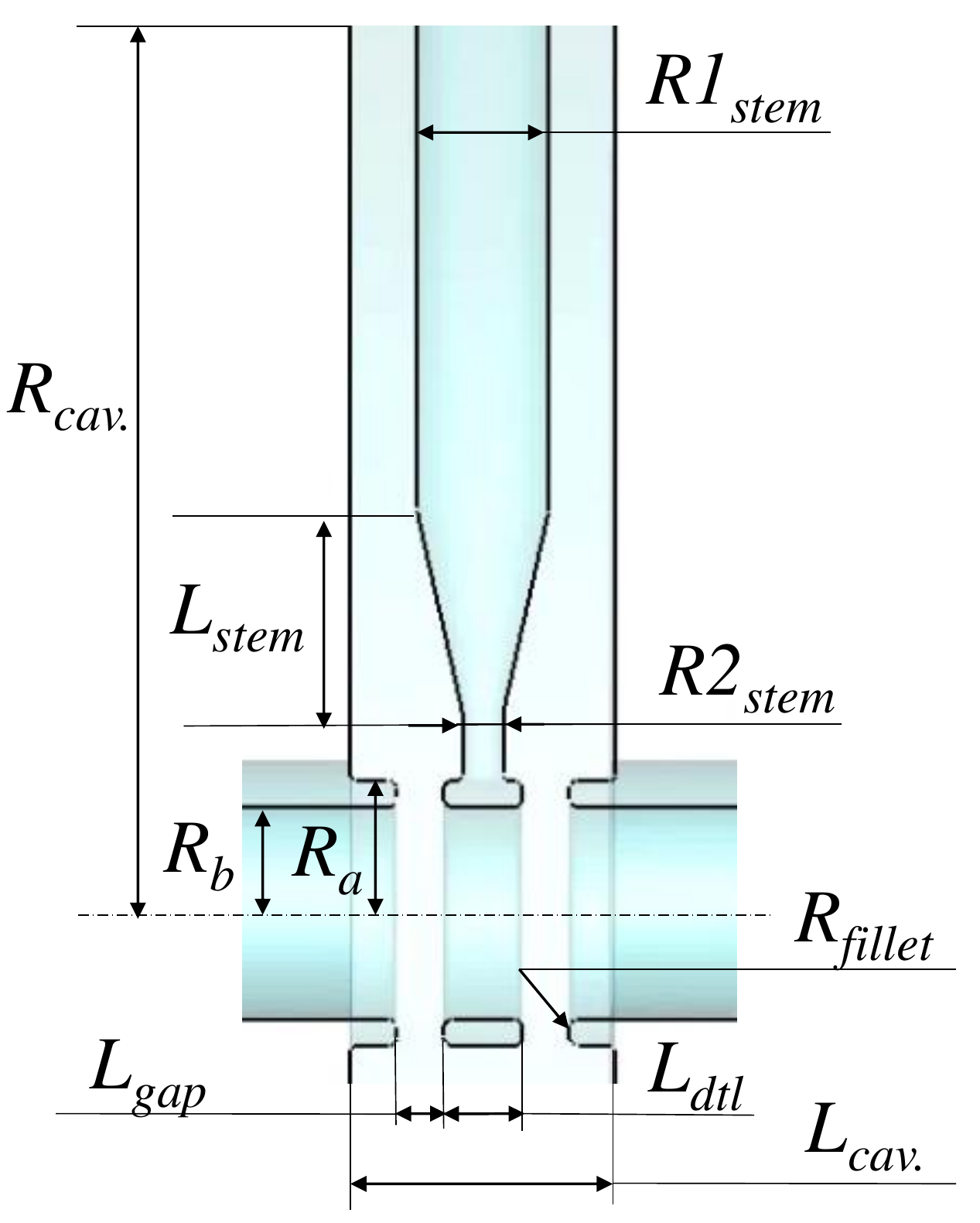}
   \end{center}
   \caption{Cutaway view of the three dimensional model of the buncher cavity. } 
   \label{fig:bun_dim}
\end{figure}

The buncher cavity is designed using CST MW Studio.  
Figure~\ref{fig:bun_dim} shows the cutaway of the three-dimensional model.  
The inner radius of the drift tube ($R_{b}$) is determined from the transverse beam size obtained from the simulations. 
The fillet radius of the drift tube ($R_{fillet}$) is adjusted to make the surface field lower and then the outer radius of the drift tube ($R_{a}$) is decided. 
The lengths of the cavity, drift-tube, and gap ($L_{cav.}$, $L_{dtl}$, and $L_{gap}$) are designed with the particle velocity to account for the limited space of the experiment. 
The stem dimensions ($R1_{stem}$, $R2_{stem}$, $L_{stem}$) are determined on the basis of mechanical strength. 
The cavity radius ($R_{cav.}$) is tuned so that the operational frequency is 324~MHz. 
The cavity dimensions are summarized in Table~\ref{tbl:dim}. 
\begin{table}[htp]
   \centering
   \caption{Cavity dimensions}
   \begin{tabular}{lr}
       \hline
       Dimensions               & Values [mm] \\
       \hline
       $R_{b}$  & 16\\
       $R_{a}$  & 20\\
       $R_{fillet}$  & 1.5\\
       $L_{cav}$  & 40\\
       $L_{dtl}$  & 12\\
       $L_{gap}$  & 6.9\\
       $R1_{stem}$  & $\phi$$15-20$ (ellipse) \\
       $R2_{stem}$  & $\phi$6\\
       $L_{stem}$  & 17.7\\
       $R_{cav}$  & 185.9\\
       \hline
   \end{tabular}
   \label{tbl:dim}
\end{table}%
\par
\par
The RF parameters obtained using CST MW Studio are summarized in Table~\ref{tbl:design_rf}. 
The power needed to supply the required voltage is 0.21~kW, which is sufficiently small and satisfies the requirement. 
\begin{table}[htp]
   \centering
   \caption{RF parameters.}
   \begin{tabular}{lr}
       \hline
       Parameters               & Values \\
       \hline
       Frequency [MHz]        & 324.01 \\
       Effective voltage [kV]  & 5.3 \\
       $Q_{0}$  & $3.08\times10^3$ \\
       $E_{pk}$ [MV/m]  & 2.8 \\
       Power dissipation [kW]  & 0.21 \\
       $R_{sh}$ [M$\mathrm{\Omega}$]  & 0.13 \\
       \hline
   \end{tabular}
   \label{tbl:design_rf}
\end{table}
\par
In a QWR structure, a dipole field exists in the acceleration gaps~\cite{bib:qwr_dipole}. 
The dipole field effect on the beam dynamics is investigated by using GPT. 
\reva{The dipole field effect is negligible and the result is consistent to that obtained from PARMILA. }
Because \reva{the effect is} negligible, conventional correction methods such as 
shifting the inner drift tube was not implemented. 
\par
In the longitudinal bunch size measurement, it is not necessary to tune the resonant frequency to 324~MHz precisely; 
it is required to set the resonant frequency to the same value as that of the RFQ. 
In addition, there is no available space for a frequency tuning system such as a conductive tuner. 
Therefore, no frequency tuning system is implemented in the buncher cavity \rev{and the operation frequency of the RFQ is tuned to that of the buncher cavity. } 
\par
\begin{figure}[htp]
   \begin{center}
    \includegraphics[width=0.45\textwidth]{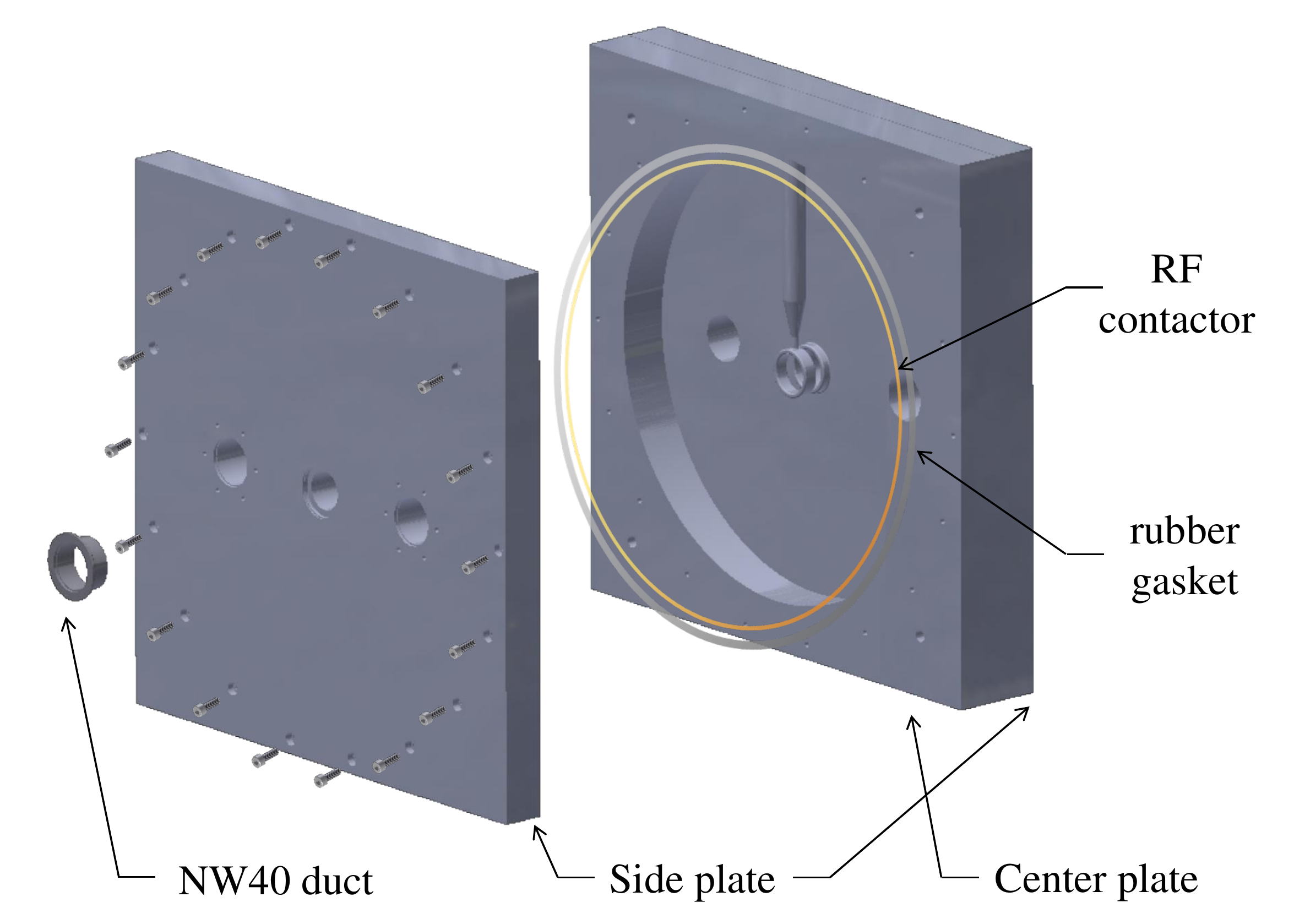}
   \end{center}
   \caption{Mechanical structure of the buncher cavity. } 
   \label{fig:3dmodel}
\end{figure}

The buncher cavity was fabricated using a three-piece design, as shown in Fig.~\ref{fig:3dmodel}. 
The two side plates are connected to the center plate via an RF contactor and a Viton O-ring. 
The drift tube and stem are machined in the center plate as a monolithic structure. 
The material of the cavity is oxygen-free copper (OFC, JIS C1020). 
The ISO KF40 duct and the side plate are attached by vacuum brazing. 
The transverse length of the buncher cavity is 450~mm, and the longitudinal length including the NW40 duct is 142~mm. 
\par
\begin{figure}[htp]
   \begin{center}
    \includegraphics[width=0.45\textwidth]{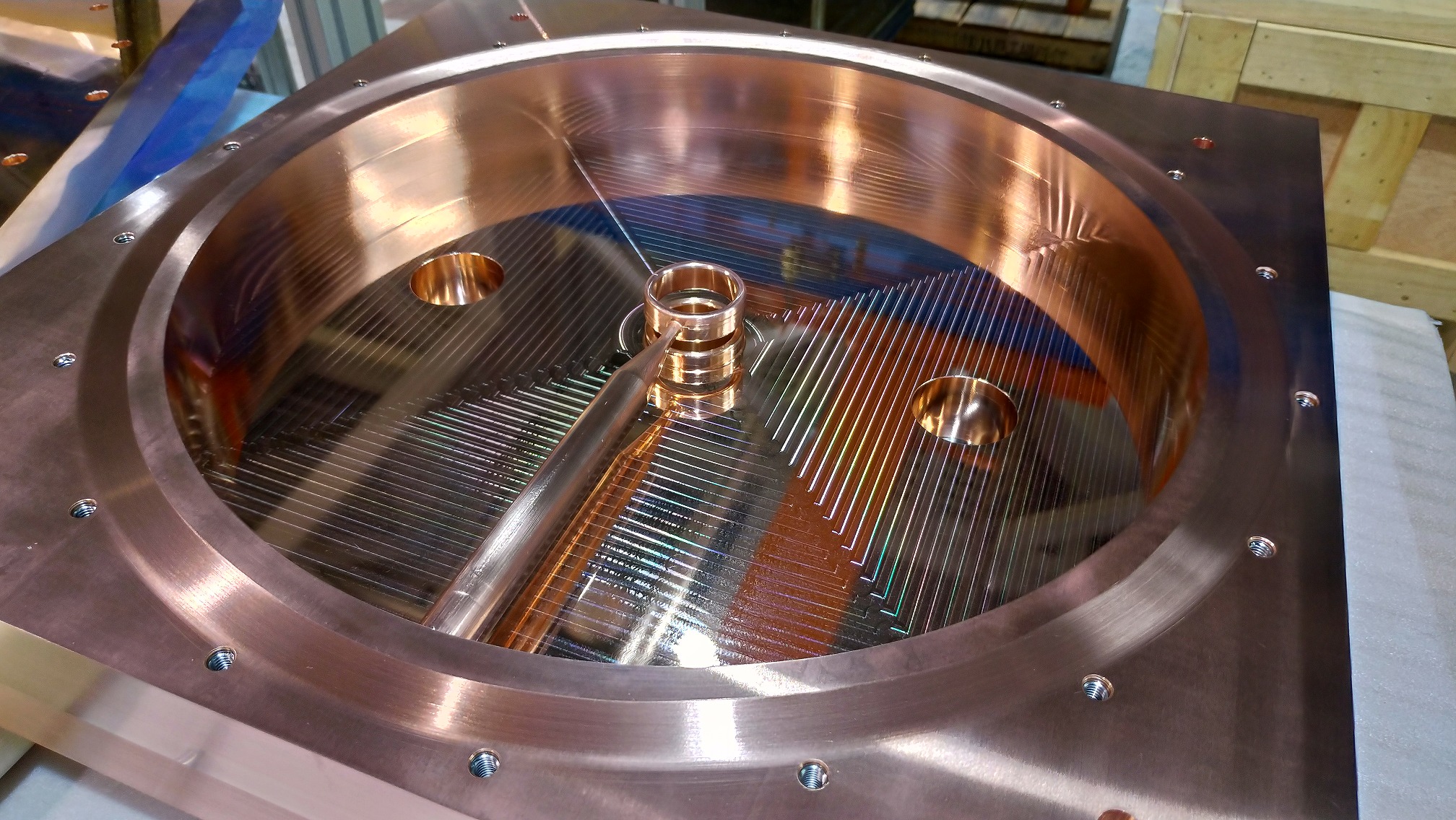}
   \end{center}
   \caption{Center plate connected to a side plate. } 
   \label{fig:bun_photo}
\end{figure}

Figure~\ref{fig:bun_photo} shows the fabricated center plate. 
Three-dimensional measurement after fabrication showed that the fabrication had an accuracy of approximately 0.05~mm. 
This fabrication error corresponds to 30~kHz. 
\rev{\section{RF measurements and results}\label{sec:result}}
Measurements of the resonant frequency $f$ and the unloaded quality factor $Q_0$ were performed using a Vector Network Analyzer (VNA). 
Table~\ref{tbl:microwave_measurement} shows the measured and simulated values of $f$ and $Q_0$. 
The measured resonant frequency is in good agreement with the simulated one. 
The discrepancy of 0.02\% between the measured and simulated frequencies is considered to be due to the effect of the loop-type of the RF pickup. 
The measured $Q_0$ is about 99\% of the simulated one. 
\begin{table}[htp]
   \centering
   \caption{Resonant frequency and quality factor of the buncher cavity.}
   \begin{tabular}{lrr}
       \hline
       Parameters        &   simulation & measurement \\
       \hline
	Resonant frequency (MHz) & 324.01  & 323.95 \\ 
	Quality factor  &  $3.08\times10^3$ & $3.06\times10^3$ \\
       \hline
   \end{tabular}
   \label{tbl:microwave_measurement}
\end{table}
\par
\begin{figure}[htp]
   \begin{center}
    \includegraphics[width=0.45\textwidth]{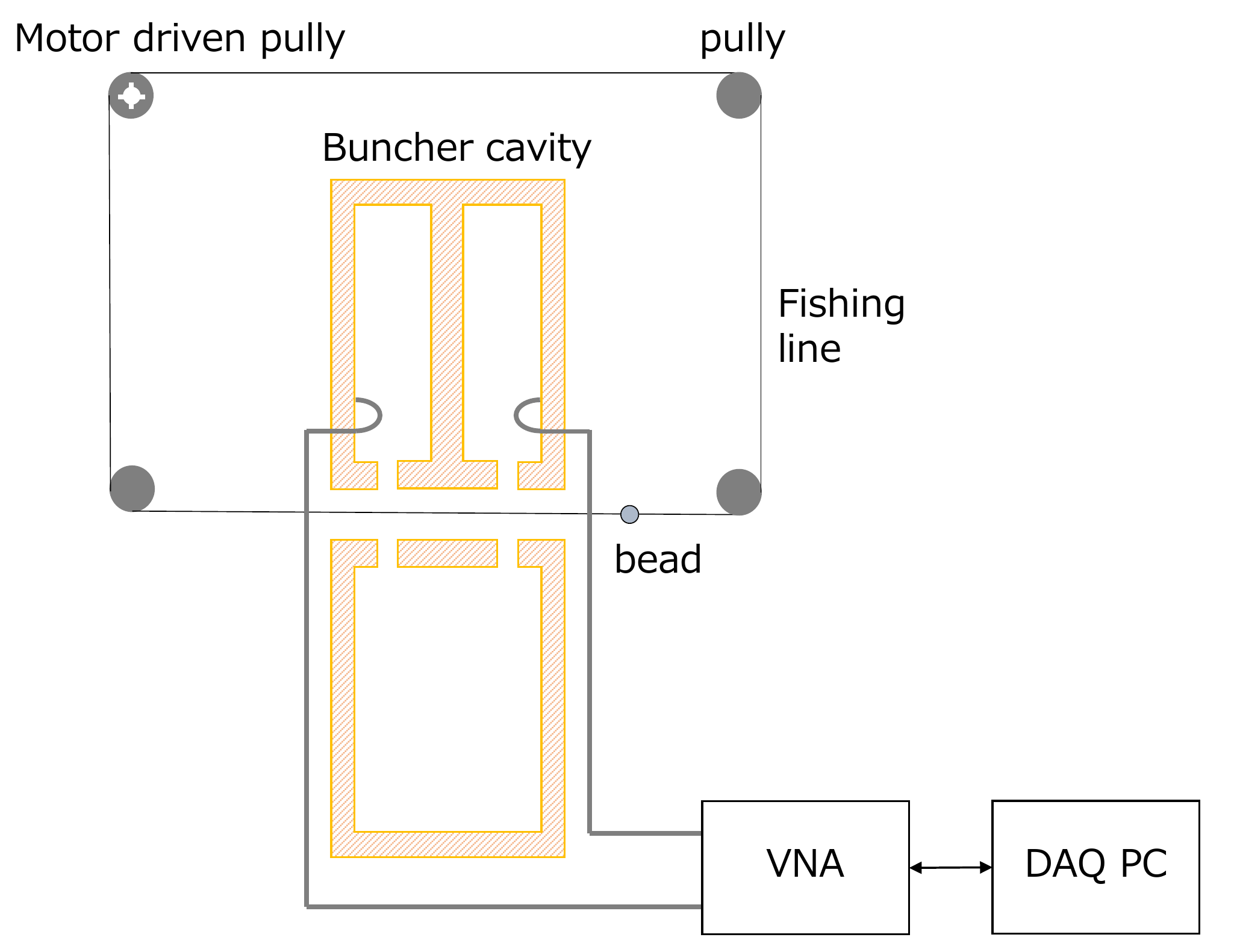}
   \end{center}
   \caption{Bead pull measurement setup} 
   \label{fig:bpm_setup}
\end{figure}
\begin{figure}[b]
   \begin{center}
    \includegraphics[width=0.5\textwidth]{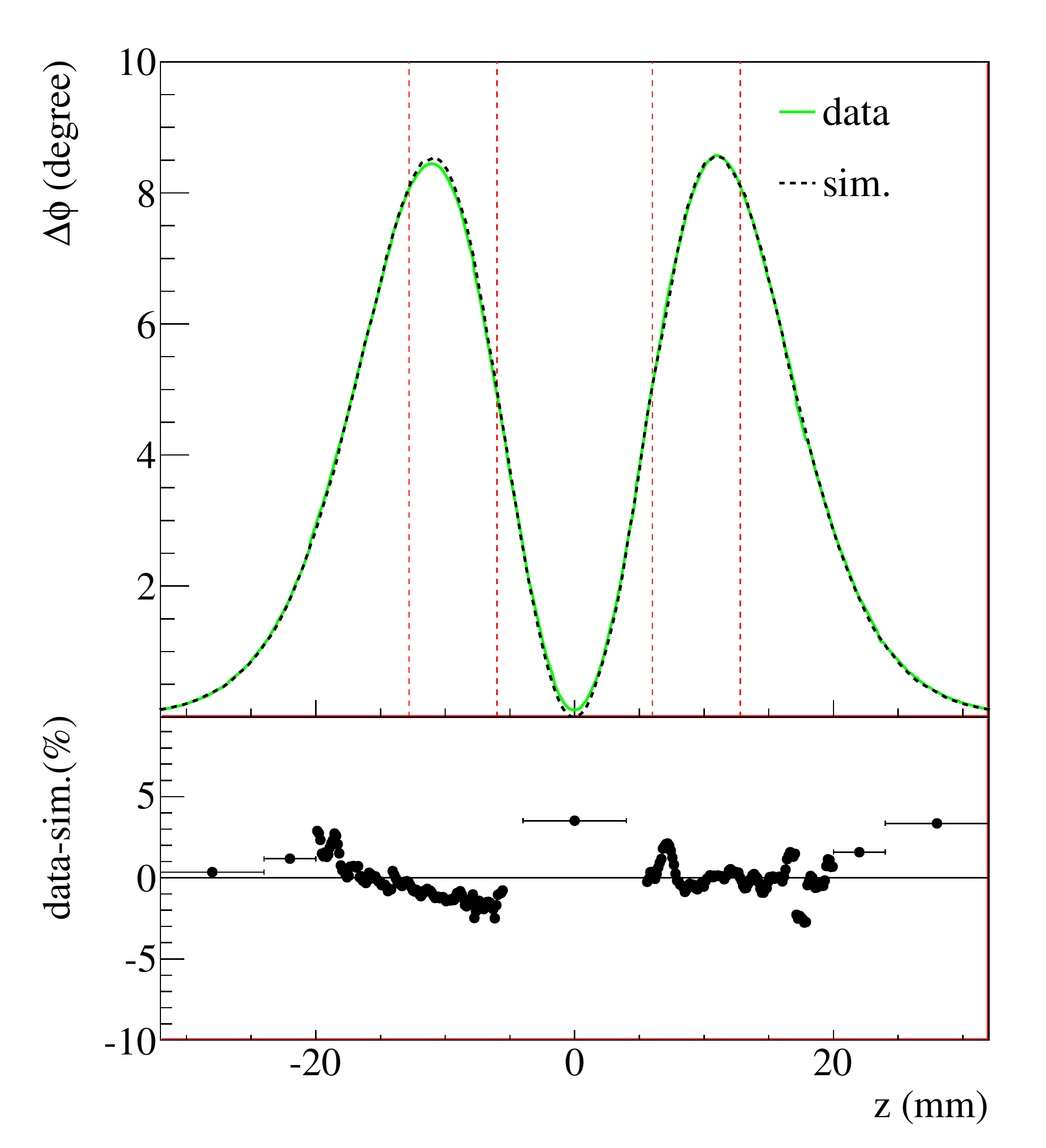}
   \end{center}
   \caption{Phase shift by the bead pull measurement. (Top) measured (green solid line) and simulated (black dotted line) phase shift. 
    The red dotted line shows the gap region. 
    (Bottom) difference between measurement and simulation.  } 
   \label{fig:result_bpm}
\end{figure}

Figure~\ref{fig:bpm_setup} shows a bead pull measurement setup~\cite{bib:bpm}. 
A 3~mm diameter spherical metal bead on a fishing line is advanced by a motor driven pully. 
The value of S21 is measured with the VNA while the metal bead is moving. 
The result of the phase shift measurement is shown in Fig.~\ref{fig:result_bpm}. 
The phase shift is proportional to $\varepsilon_{0} E^2 - \mu_{0} H^2/2$~\cite{bib:cernrf}, where $\varepsilon_{0}$ is the permittivity, $E$ is the electric field, 
$\mu_{0}$ is the magnetic permeability, and $H$ is the magnetic field. 
Two phase-shifting cycles are observed due to the double-gap. 
The measured phase shift along the z-direction is in excellent agreement with the simulated one. 
Especially around the gap, the difference is less than 4\%, which is within the uncertainties of measurement due to 
the fishing line alignment and accuracy of the phase shift. 
\par
\par
The longitudinal bunch size measurement was performed for six days in November 2018. 
A loop-type RF coupler was inserted into the side plate. 
The RF pulse was generated by a Tektronix signal generator TSG4104~\cite{bib:tektronix}, 
and then amplified by a 324-MHz 5~kW solid-state amplifier unit, R\&K CA324BW0.4-6767R(P)~\cite{bib:amp}. 
The RF power was transmitted via 50-$\Omega$ coaxial cables. 
The RF pulse width was 100~$\mu$s, and the repetition rate was 25~Hz; these were the same as that of the muon beam. 
The relative phase of the RFQ was tuned by a trombone phase shifter with an accuracy of less than 1~degree. 
The RF power and phase were monitored using a loop pickup monitor that was inserted into the cavity. 
\par
During measurement, the buncher cavity was successfully operated. 
The relative phase shift between the buncher cavity and the RFQ was stable within 1~degree. 
The pick-up power was stable within a few percent during the experiment. 
\par
\section{Conclusion}\label{sec:conclusion}
A buncher cavity has been developed for the bunch size measurement after muon acceleration by the RFQ. 
It is designed for $\beta=0.04$ with a frequency of 324~MHz. 
It employs a double-gap structure operated in the TEM mode to account for the limited space of the experiment. 
\par
The buncher cavity was designed using CST WM Studio. 
The cavity inner radius and total length are 185.9~mm and 142~mm, respectively. 
It supplies an effective voltage of 5.3~kV with 0.21~kW for longitudinal bunch size measurement with compact dimensions. 
\par
Microwave and bead pull measurements were performed after fabrication. 
The resonant frequency was found to be in good agreement within 0.02\%. 
The measured $Q_0$ was about 99\% of the simulated one. 
\par
The buncher cavity was successfully operated for the longitudinal bunch size measurement of muons accelerated by the RFQ. 
\begin{acknowledgments}
We express our appreciation to TOTAL INTEGRATOR MACHINERY\&ENGINEERING Co., who fabricated the buncher cavity.
This work is supported by JSPS KAKENHI Grant Numbers
JP16H03987, 18H05226, and JP18H03707. 
The experiment at the Materials and Life Science Experimental Facility
of J-PARC was performed under user programs (Proposal No. 2018A0222).
\end{acknowledgments}


\end{document}